%% file: Bs-Jpsif0.tex
\RequirePackage{lineno}
\documentclass[aps,prl,twocolumn,showpacs,superscriptaddress,groupedaddress]{revtex4}  
\usepackage{graphicx}  
\usepackage{dcolumn}   
\usepackage{bm}        
\usepackage{amssymb}   
\usepackage{amsmath}
\usepackage{epstopdf}

\graphicspath{{./graficas/}}

\begin{document}
 
\title{$B^{0}_{s}$ lifetime measurement in the CP-odd decay channel $B^{0}_{s} \to J/\psi\mbox{ }f_{0}(980)$}

\input author_list.tex

\date{\today}

\begin{abstract}

The lifetime of the $B^{0}_{s}$ meson is measured in the decay channel 
$B^{0}_{s} \to J/\psi\mbox{ }\pi^+ \pi^-$ with $880 \leq M_{\pi^+\pi^-} \leq 1080$ MeV/$c^2$, 
which is mainly a CP-odd state and dominated by the $f_{0}(980)$ 
resonance. In 10.4 fb$^{-1}$ of data collected with the D0 detector in 
Run II of the Tevatron, the lifetime of the $B^{0}_{s}$ meson is measured to 
be $\tau(B^{0}_{s}) = 1.70\pm 0.14 \mbox{ (stat)} \pm 0.05 \mbox{ (syst) 
ps}$. Neglecting CP violation in $B_s^0/\bar{B}_s^0$ mixing, the 
measurement can be 
translated into the width of the heavy mass eigenstate of the $B^0_s$, 
$\Gamma_H = 0.59 \pm 0.05 \mbox{ (stat)} \pm 0.02 \mbox{ (syst)} \mbox{ 
ps$^{-1}$}$.
 \end{abstract}

\maketitle

\vskip.5cm

The $B_s^0$ and $\bar{B}_s^0$ mesons are produced as flavor eigenstates  
at hadron colliders, but the particles propagate as mass eigenstates. 
There are two mass eigenstates, the so-called heavy and light states, 
which are linear combinations of the flavor eigenstates. In the absence of 
CP-violation in mixing, the mass eigenstates are also CP eigenstates, with the 
heavier state expected to be the CP-odd state. The lifetimes of the two mass 
eigenstates can be different from each other and different from the average $B_s^0$ 
lifetime. A measurement of the $B_s^0$ lifetime in either a pure CP-odd state or pure 
CP-even state would give important additional information about the $B_s^0$ system.

The $B_{s}^{0} \to J/\psi f_{0}(980)$ decay channel corresponds to a pure CP-odd eigenstate decay due 
to angular momentum conservation, since the parent $B_s^0$ is spin 0, the $f_0(980)$ 
has $J^{PC} = 0^{++}$, and the $J/\psi$ has $J^{PC}=1^{--}$. Throughout this Letter, 
the appearance of a specific charge state also implies its charge conjugate. This decay channel was 
first observed by the LHCb collaboration~\cite{lhcb}, and later 
confirmed by the Belle~\cite{bell}, CDF~\cite{cdf} and D0 ~\cite{d0} collaborations. A measurement 
of the $B_s^0$ lifetime in this channel gives access to the lifetime of the heavy 
mass eigenstate. The lifetime measurement can be transformed into a measurement of 
the parameter $\Gamma_{H}$, the decay width of the heavy $B_{s}^{0}$ mass eigenstate.
 CDF~\cite{cdf} and LHCb~\cite{lhcblifetime} have measured this lifetime, reporting 
$\tau(B^{0}_{s}) = (1.70 \pm 0.12 \pm 0.03) \mbox{ ps}$ and $\tau(B^{0}_{s}) = (1.70 
\pm 0.04 \pm 0.026) \mbox{ ps}$ respectively, which are in good agreement with each 
other and somewhat longer than the mean lifetime $\tau(B^{0}_{s}) = (1.52 \pm 0.007) \mbox{ ps}$
\cite{PDG2014}.

In this analysis, we report the lifetime of the $B^{0}_{s}$ meson measured in the 
decay channel $B^{0}_{s} \to J/\psi(\to \mu^+\mu-)\pi^+ \pi^{-}$ with 
$880 \leq M_{\pi^+\pi^-} \leq 1080$ MeV/$c^2$, which is dominated by the $f_{0}(980)$ 
resonance and which is CP-odd at the ~99\% level \cite{lhcb1,lhcb2}. 
The data used in this analysis were collected with the 
D0 detector during Run II of the Tevatron collider at a center-of-mass energy of 
1.96~TeV, and correspond to an integrated luminosity of 10.4~fb$^{-1}$.

The D0 detector is described in detail elsewhere~\cite{run2det}. The detector 
components most relevant to this analysis are the central tracking and the muon 
systems. The former consists of a silicon microstrip tracker (SMT) and a central 
scintillating fiber tracker (CFT) surrounded by a 2~T superconducting solenoidal 
magnet. The SMT has a design optimized for tracking and vertexing for pseudorapidity 
of $|\eta|<3$~\cite{etadef}. For charged particles, the resolution on the distance of 
closest approach as provided by the tracking system is approximately 50~$\mu$m for 
tracks with $p_T \approx 1$~GeV/$c$, where $p_T$ is the component of the momentum 
perpendicular to the beam axis. It improves asymptotically to 15~$\mu$m for tracks 
with $p_T > 10$~GeV/$c$. Preshower detectors and electromagnetic and hadronic 
calorimeters surround the tracker. The muon system is located outside the calorimeter, 
and consists of multilayer drift chambers and scintillation counters inside 1.8~T 
iron toroidal magnets, and two similar layers outside the toroids. Muon identification and tracking for 
$|\eta|<1$ relies on 10 cm wide drift tubes, while 1 cm mini-drift tubes are used for 
$1<|\eta|<2$. 
We base our data selection on reconstructed charged tracks and identified
muons. 
Events used in this analysis are collected with both single muon and dimuon triggers.
To avoid a trigger bias in the lifetime  measurement, we reject events that satisfy only impact parameter-based triggers.
We simulate signal events with
PYTHIA  \cite{pythia} and EvtGen \cite{evtgen}, followed by full detector simulation using GEANT3 \cite{geant3}.  
To correct for trigger effects, we weight simulated events so that the $p_T$ distributions of the muons match the distributions in data.

The $B^0_{s}$ reconstruction begins by reconstructing $J/\psi$ candidates 
followed by searching for $\pi^+\pi^{-}$ candidates.
To reconstruct 
$J/\psi\to\mu^{+}\mu^{-}$ candidates, events with at least two muons of 
opposite charge reconstructed in the tracker and the muon system are 
selected. For at least one of the muons, hits are required in the muon system both inside and outside of the toroids.
Both muons must have hits in the SMT and have 
$p_T>$2.5~GeV/$c$. The muon tracks are constrained to originate from a common 
vertex with a $\chi^2$ probability greater than 1\%. Each $J/\psi$ candidate 
is required to have a $p_T$ greater than 1.5 GeV/$c$ and a mass in the range 
2.80--3.35~GeV/$c^2$. 

We require two oppositely charged tracks, assumed to have the pion mass, each with
at least two SMT hits
and at least two CFT hits, and at least eight total hits in the tracking system. 
These two tracks are  
constrained to a common vertex with a $\chi^2$ probability greater than 1\%. 
Each $\pi^+ \pi^{-}$ candidate is required to have a mass in the range 
$880 \leq M_{\pi^+\pi^-} \leq 1080$ MeV/$c^2$ and a $p_T$ greater than 1.5 GeV/$c$.  The $B^{0}_{s}$ 
candidates are reconstructed by performing a constrained fit to a common 
vertex for the two pions and the two muon tracks, with the latter constrained 
to the $J/\psi$ mass of 3.097~GeV/$c^2$~\cite{PDG2014}. The $B^{0}_{s}$ 
candidates are required to have a mass within the range 5.1--5.8~GeV/$c^2$, 
and to have a $p_T$ greater than 6.0 GeV/$c$.

To determine the decay time of the $B^{0}_{s}$, the distance traveled by the 
candidate projected in a plane transverse to the beam direction is measured, 
and then a correction for the Lorentz boost is applied. The transverse decay 
length is defined as $L_{xy} = \bm{L}_{xy} \cdot \bm{p}_T/p_T$, where 
$\bm{L}_{xy}$ is the vector that points from the primary vertex \cite {pv}
to the $B^{0}_{s}$ decay vertex, and $\bm{p}_T$ is the transverse 
momentum vector of the $B^{0}_{s}$ candidate. The event-by-event value of the proper 
transverse decay length, $\lambda$, for the $B^{0}_{s}$ candidate is given by:
\begin{equation} 
    \label{ctau} \lambda = L_{xy} \frac{cM_{B}}{p_{T}}, 
\end{equation} 
\noindent 
where $M_{B}$ is the world average mass value of the $B^{0}_{s}$ meson~\cite{PDG2014}.
In order to remove background, $B^{0}_{s}$ 
candidates are 
required to have $\lambda>$ 0.02 cm and 
uncertainties on $\lambda$ of less than 0.01 cm.

A simultaneous unbinned maximum likelihood fit to the mass and proper decay 
length distributions is performed to measure the lifetime.
The likelihood function ${\cal{L}}$ is defined by:
\begin{equation}
        {\cal{L}} =
        \prod^{N}_{j=1} \left[N_{\text{sig}}{\cal{F}}_{{\text{sig}}}^{j} + N_{\text{comb}}{\cal{F}}_{{\text{comb}}}^{j} + N_{\text{xf}}{\cal{F}}_{{\text{xf}}}^{j} + N_{B^+}{\cal{F}}_{{B^+}}^{j} 
\right],
\end{equation}
 where $N$ is the total number of events and $N_{\text{sig}}$, $N_{\text{comb}}$, $N_{\text{xf}}$ and $N_{B^+}$ are the expected number of signal, 
combinatorial background, cross-feed contamination and 
$B^{\pm} \to J/\psi K^{\pm}$ 
 events in the sample, respectively. All these parameters are determined in the fit. The different background contributions are discussed below. 

The functions ${\cal{F}}$ are the product of three probability density 
functions that model distributions of the mass 
 $m$, the proper transverse decay length $\lambda$, and the uncertainty on the proper decay length $\sigma_\lambda$ for the signal, combinatorial background, cross-feed contamination, and $B^{\pm}$ events 
\begin{multline}
\mathcal{F}^j_{\alpha} = M_{\alpha}(m_j) T_{\alpha}(\lambda_j|\sigma_{\lambda_j}) E_{\alpha}(\sigma_{\lambda_j}); \\
\hspace{0.2cm} \alpha = \{\text{sig, comb, xf, }B^+\},
\end{multline}
where $m_j$, $\lambda_j$, and $\sigma_{\lambda_j}$ represent the mass, the transverse proper decay length, and its uncertainty, respectively, for a given event 
$j$. 
The use of the probability density functions $T$ and $E$ follows the method of reference~\cite{punzi}.
The specific models and parameters used in the fit are described below. 

For the signal, the mass distribution is modeled by a Gaussian function
, $M_{\text{sig}}(m_j) = G(m_j; \mu_m, \sigma_m)$, where
\begin{equation}\label{eq:massgauss}
 G(m_j; \mu_m, \sigma_m)=\frac{1}{\sqrt{2\pi}\sigma_m} e^{-(m_j-\mu_m)^2/(2\sigma_m^2)} ,
\end{equation} 
with $\mu_m$ and $\sigma_m$ the mean and the width of the Gaussian, determined from the fit.

The combinatorial background is primarily due to random combinations of $J/\psi$'s with additional tracks in the event, and its mass distribution is described by an exponential function 
\begin{equation}    
M_{\text{comb}}(m_j;a_0) = e^{a_0 m_j},
\end{equation}
 with $a_0$ determined from the likelihood fit. 

The 
physics cross-feed contamination is mainly produced by the combination of $J/\psi$ mesons from $b$ hadron decays with other 
particles produced in the collision, including from the same $b$ hadron.  Other $b$ hadron decays with final states such as 
$B^{0}\to J/\psi K \pi$, $B^{0}\to J/\psi \pi \pi$ and $B^0_s\to J/\psi K K$ are reconstructed at mass below the signal of 
the $B^0_s$, either due to the lower mass of the $B^0$ 
or the incorrect mass assignment of the pion mass to a kaon track.
Simulations of these decays show that the cross-feed contamination can be described by a single Gaussian component
\begin{equation}
M_{\text{xf}}(m_j) = G(m_j;\mu_{\text{xf}},\sigma_{\text{xf}}),
\end{equation}
where $\mu_{\text{xf}}$ and $\sigma_{\text{xf}}$ are the mean and the width of the Gaussian, determined from the likelihood fit.

The final contribution arises from $B^{\pm} \to J/\psi K^{\pm}$ 
decays in which the kaon has been assigned a pion mass, and an additional 
track accidentally forms a vertex with the $J/\psi K^{\pm}$. 
The candidate  mass is reconstructed in the region of real $B^0_s$ events.
If the higher $p_T$ non-muon track in $B_s^0$ 
candidates is assigned a kaon mass, a clear $B^{\pm}$ signal emerges. Events 
in this $B^{\pm}$ mass peak, when interpreted as $J/\psi \pi \pi$, are used 
as a template~\cite{cranmer} to determine the shape of the mass distribution of the $B^{\pm} 
\to J/\psi K^{\pm}$ contamination in the $B_s^0$ candidates.

The $\lambda$ distribution for the signal is parameterized by an exponential decay
convoluted with 
a resolution function
\begin{equation}
    T_{\text{sig}}(\lambda_{j} | \sigma_{\lambda_j})=\frac{1}{\lambda_{B}}\int^{\infty}_{0}G(x;\lambda_{j},\sigma_{\lambda_j})\exp 
\left(\frac{-x}{\lambda_{B}}\right) dx,
\end{equation}
with $\lambda_{B} = c\tau$ of the $B^{0}_s$ to be measured. 
The $\lambda$ distribution for the background components is parametrized by the sum of two exponential decay functions 
modeling combinatorial background $T_{\text{comb}}(\lambda_{j})$, an exponential decay for the cross-feed contamination $T_{\text{\text{xf}}}(\lambda_{j})$, and an exponential decay 
function that describes $T_{B^+}(\lambda_{j})$ for $B^{\pm}$ contamination. 

The distribution of the $\lambda$ uncertainty $E_{\text{sig}}(\sigma_{\lambda_j})$ is described by a phenomenological model, using an exponential with decay constant $1/\zeta$, convoluted 
with a Gaussian with mean $\epsilon$ and width $\delta$: 
\begin{equation} 
E_{\text{sig}}(\sigma_{\lambda_j};\zeta,\epsilon,\delta)= \frac{1}{\zeta} e^{ -\sigma_{\lambda_j}/\zeta} \otimes 
G(\sigma_{\lambda_j};\epsilon,\delta),
\end{equation}
where the parameters $\zeta$, $\epsilon$ and width $\delta$ are determined from the fit in the sample of events. The uncertainties in $\lambda$ for the background components are treated in the same manner. 

\begin{figure}
        \includegraphics[scale=.45]{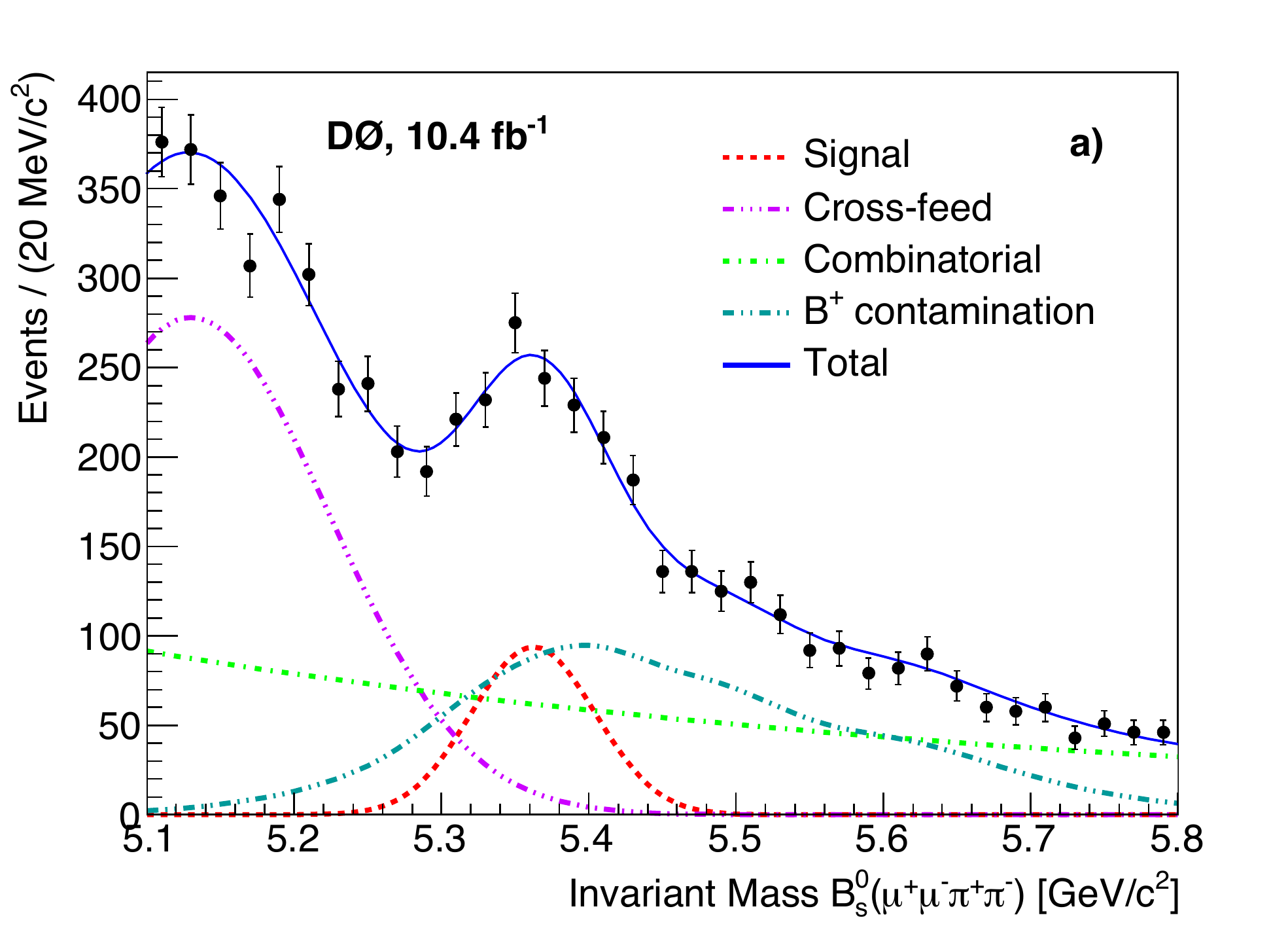} \\ 
        \includegraphics[scale=.45]{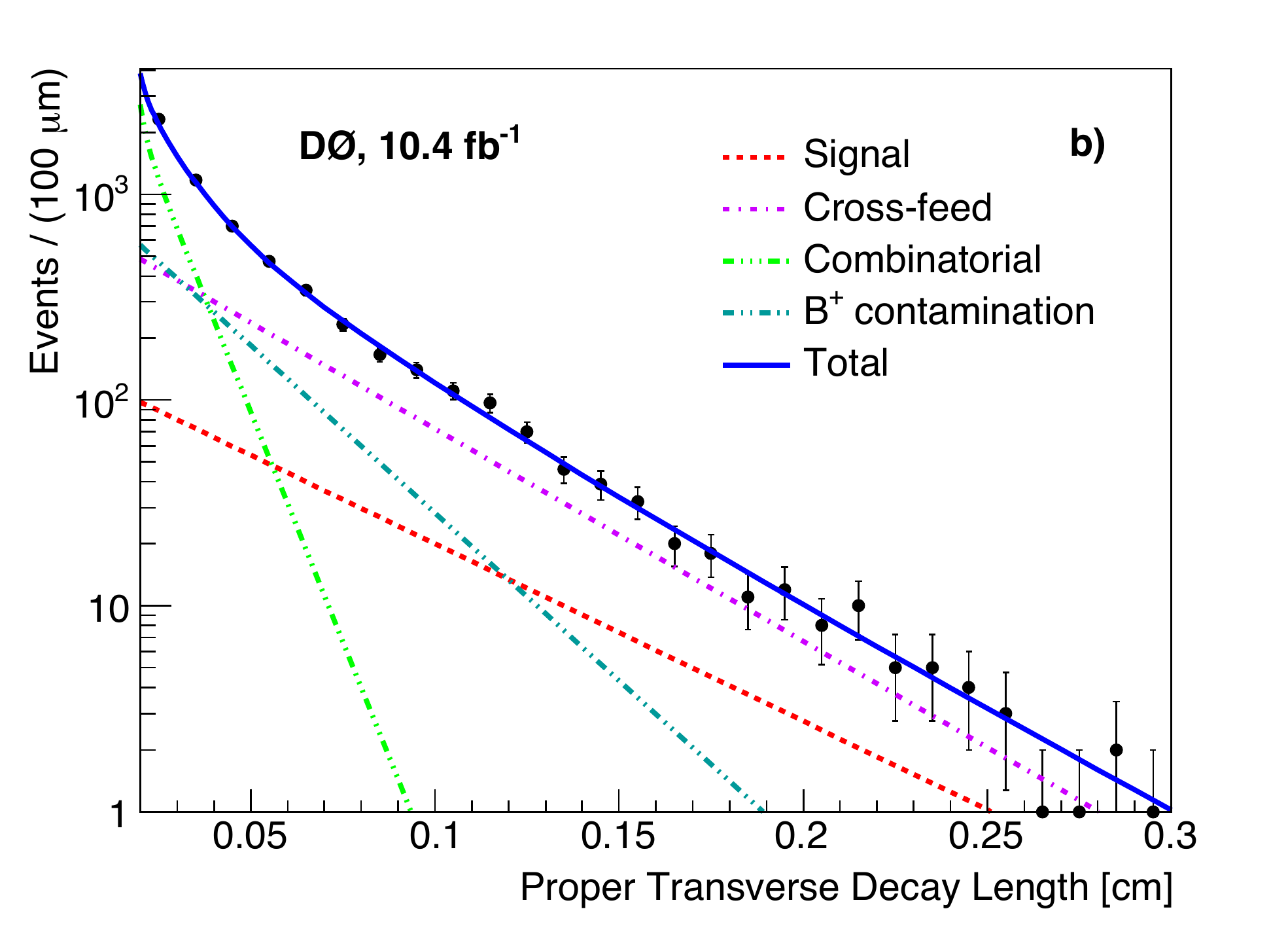} \\
        \includegraphics [scale=.45]{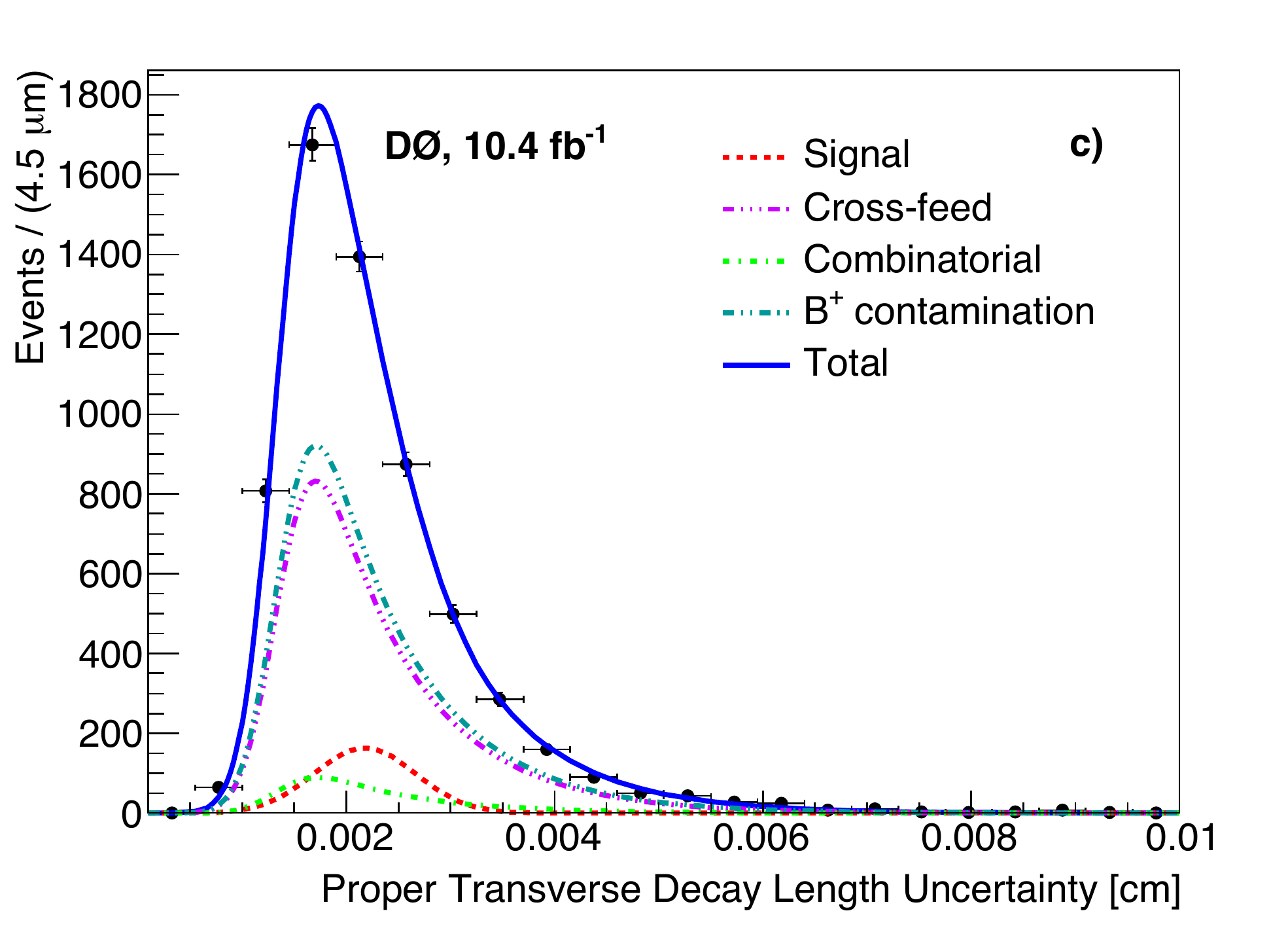}
    \caption{\label{fig:lbm} Distributions of (a) invariant mass, (b) proper transverse decay length, and (c) proper transverse decay length uncertainty for $B^0_{s}$ candidates, 
      with the fit results superimposed.  Each of the different background components is indicated in the figure. 
 The fit yields $c\tau (B^{0}_{s}) = 504\pm 42$ $~\mu$m.}
\end{figure}

\begin{figure}
\includegraphics [scale=.45]{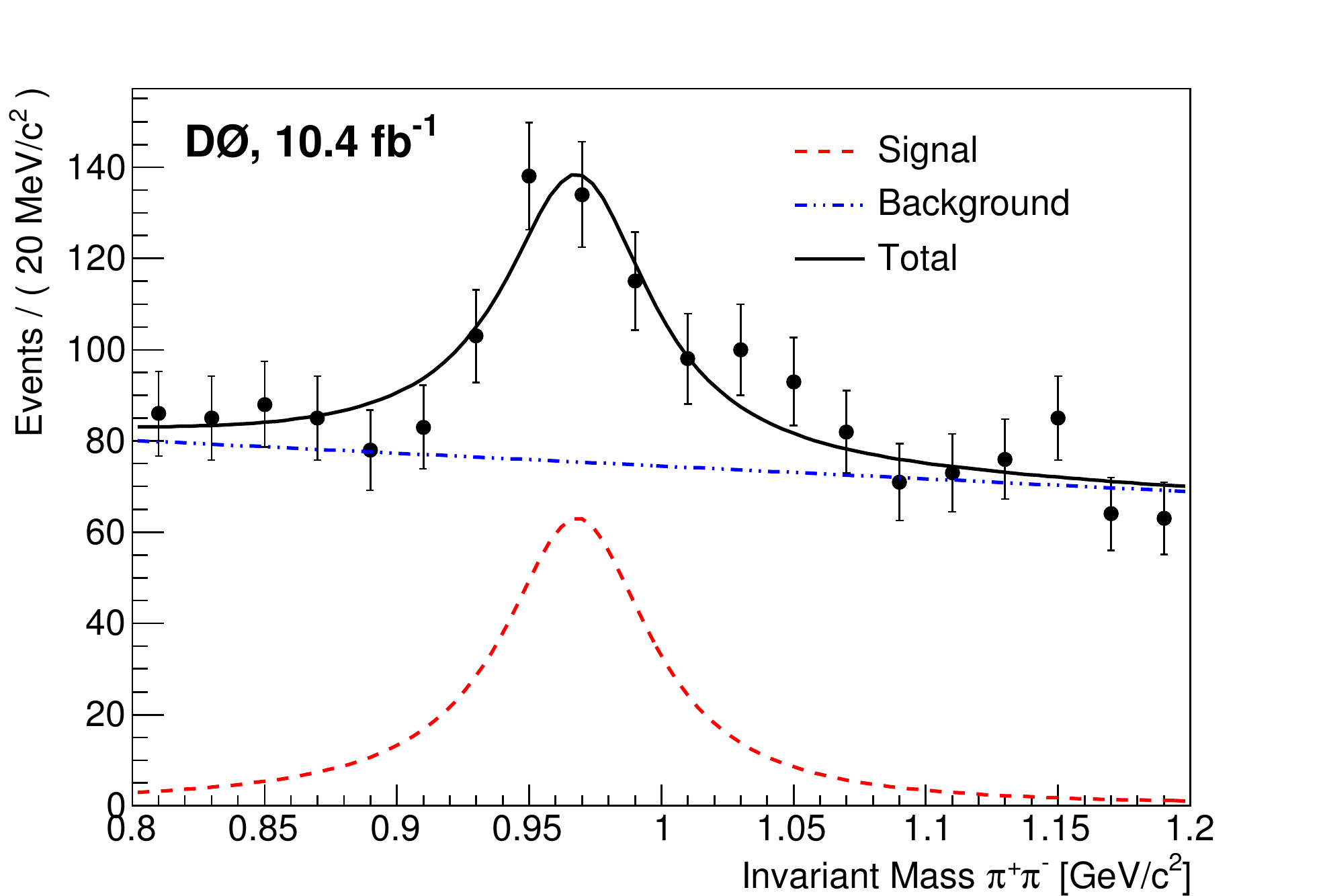}
\caption {$M(\pi^+\pi^-)$ distribution for events with $M(\mu^+\mu^-\pi^+\pi^-)$ within $\pm 1 \sigma$ of the $B_s^0$ mass. }
\label {f0}
\end{figure}

 The fit yields $c\tau (B^{0}_{s}) = 504\pm 42$ $~\mu$m and the numbers of signal decays to be $ 494\pm 85$. 
Figure~\ref{fig:lbm} shows the mass, $\lambda$ and $\lambda$ uncertainty distributions for data with the fit results superimposed.
Figure~\ref{f0} shows the $M(\pi^+\pi^-)$ mass distribution for events with $M(\mu^+\mu^-\pi^+\pi^-)$ within one $\sigma$ of the $B_s^0$ 
mass. The $M(\pi^+\pi^-)$ distribution is fit with a Flatt\' e function \cite{flatte,flatte2,flatte3} and a polynomial background.   

Table~\ref{tab:table2} summarizes the systematic uncertainties considered for 
this measurement. The contribution from possible misalignment of the SMT 
detector has been previously determined to be 5.4~$\mu$m~\cite{PRL94-102001}. 
The invariant mass window used for the $\pi^+ \pi^-$ distribution 
is varied from its nominal value of 200 MeV/$c^2$ to 160 and 240 MeV/$c^2$ and the fit
is performed for each new mass window 
selection. This results in a systematic uncertainty of 8 $\mu$m. 
We test the modeling and fitting method used to estimate the lifetime using data generated in pseudoexperiments with a range of lifetimes from 300 to 800 $\mu$m. A bias arises due to imperfect separation of signal and background. Since the background has a shorter lifetime than the signal, the result is a slight underestimate of the signal lifetime. The bias has a value of -4.4 $\mu$m for an input lifetime of 500 $\mu$m and 500 signal events. We have corrected the lifetime for this bias and a 100\% uncertainty on the correction has been applied to the result.
We estimate the systematic uncertainty due to the models for the $\lambda$ and mass distributions by varying the parameterizations of the different components: (i) the cross-feed contamination is modeled by two Gaussian functions instead of one, (ii) the exponential mass distribution for the combinatorial background model is replaced by a first order polynomial, (iii) the smoothing of the non-parametric function that models the $B^\pm$ contamination is varied, and (iv) 
the exponential functions modelling the background $\lambda$ distributions are smeared with a Gaussian resolution similar to the signal.
To take into account correlations between the effects of the different models, a fit that combines all different model changes is performed. We quote the difference between the result of this fit and the nominal fit as the systematic uncertainty.

Several cross-checks of the lifetime measurement are performed. The mass 
windows are varied, the reconstructed $B^0_s$ mass is used instead of the 
world average~\cite{PDG2014} value, and the data sample is split into 
different regions of pseudorapidity and of azimuthal angle.  All results 
obtained with these variations are consistent with the nominal measurement. 
Using the $B^{\pm}$ background sample extracted from the data, we performed a fit for the lifetime of this component of the background. The result is in good agreement with the values obtained from the global fit. We have also fit the lifetime of the cross-feed contamination from the simulation and again good agreement with the global fit is observed.

In order to estimate the effect of a small non CP-odd component in the analysis, we performed the fit with two exponential decay components for the signal, with the lifetime of one of them fixed to the world average of the CP-even $B^0_s$ lifetime~\cite{PDG2014}, and its fraction to be 0.01 as found by the LHCb experiment~\cite{lhcblifetime}. The lifetime fit finds a variation of 1 $\mu$m with respect to the nominal fit result.

\begin{table}
\caption{\label{tab:table2} Summary of systematic uncertainties in the $B^0_s$ lifetime measurement. The
total uncertainty is determined by combining individual uncertainties in quadrature.}
\begin{ruledtabular}
\begin{tabular}{lc}
Source & Variation ($\mu$m) \\
\hline
Alignment             & 5.4  \\
$\pi^+\pi^-$ invariant mass window  &  8.0 \\
Fit bias  & 4.4 \\
Distribution models & 12.5 \\
\hline
{Total}               & 16.4 \\
\end{tabular}
\end{ruledtabular}
\end{table}

In summary, the lifetime of the $B^{0}_{s}$ is measured to be:
\begin{equation}
c\tau(B^{0}_{s}) = 508 \pm 42 \mbox{ (stat)} \pm 16 \mbox{ (syst) } \mu\mbox{m,} 
\end{equation}
\noindent from which we determine:
\begin{equation}
\tau(B^{0}_{s}) = 1.70\pm 0.14 \mbox{ (stat)} \pm 0.05 \mbox{ (syst) ps,}
\end{equation}
in the decay channel $B^0_s \to J/\psi \pi^+ \pi^{-}$ with $880 \leq M_{\pi^+\pi^-} \leq 1080$ MeV/$c^2$.
In the absence of CP violation in mixing,
this measurement
can be translated into the width of the heavy mass eigenstate of the $B^0_s$:
\begin{equation} 
\Gamma_H = 0.59 \pm 0.05 \mbox{ (stat)} \pm 0.02 \mbox{ (syst)} \mbox{ ps$^{-1}$}.
\end{equation}
This result is in good agreement with previous measurements and provides an independent confirmation of the 
longer lifetime for the CP-odd eigenstate of the $B_s^0/\bar{B}_s^0$ system. 

\input acknowledgement_APS_full_names.tex
\end{document}

%% file: author_list.tex
\affiliation{LAFEX, Centro Brasileiro de Pesquisas F\'{i}sicas, Rio de Janeiro, RJ 22290, Brazil}
\affiliation{Universidade do Estado do Rio de Janeiro, Rio de Janeiro, RJ 20550, Brazil}
\affiliation{Universidade Federal do ABC, Santo Andr\'e, SP 09210, Brazil}
\affiliation{University of Science and Technology of China, Hefei 230026, People's Republic of China}
\affiliation{Universidad de los Andes, Bogot\'a, 111711, Colombia}
\affiliation{Charles University, Faculty of Mathematics and Physics, Center for Particle Physics, 116 36 Prague 1, Czech Republic}
\affiliation{Czech Technical University in Prague, 116 36 Prague 6, Czech Republic}
\affiliation{Institute of Physics, Academy of Sciences of the Czech Republic, 182 21 Prague, Czech Republic}
\affiliation{Universidad San Francisco de Quito, Quito, Ecuador}
\affiliation{LPC, Universit\'e Blaise Pascal, CNRS/IN2P3, Clermont, F-63178 Aubi\`ere Cedex, France}
\affiliation{LPSC, Universit\'e Joseph Fourier Grenoble 1, CNRS/IN2P3, Institut National Polytechnique de Grenoble, F-38026 Grenoble Cedex, France}
\affiliation{CPPM, Aix-Marseille Universit\'e, CNRS/IN2P3, F-13288 Marseille Cedex 09, France}
\affiliation{LAL, Univ. Paris-Sud, CNRS/IN2P3, Universit\'e Paris-Saclay, F-91898 Orsay Cedex, France}
\affiliation{LPNHE, Universit\'es Paris VI and VII, CNRS/IN2P3, F-75005 Paris, France}
\affiliation{CEA Saclay, Irfu, SPP, F-91191 Gif-Sur-Yvette Cedex, France}
\affiliation{IPHC, Universit\'e de Strasbourg, CNRS/IN2P3, F-67037 Strasbourg, France}
\affiliation{IPNL, Universit\'e Lyon 1, CNRS/IN2P3, F-69622 Villeurbanne Cedex, France and Universit\'e de Lyon, F-69361 Lyon CEDEX 07, France}
\affiliation{III. Physikalisches Institut A, RWTH Aachen University, 52056 Aachen, Germany}
\affiliation{Physikalisches Institut, Universit\"at Freiburg, 79085 Freiburg, Germany}
\affiliation{II. Physikalisches Institut, Georg-August-Universit\"at G\"ottingen, 37073 G\"ottingen, Germany}
\affiliation{Institut f\"ur Physik, Universit\"at Mainz, 55099 Mainz, Germany}
\affiliation{Ludwig-Maximilians-Universit\"at M\"unchen, 80539 M\"unchen, Germany}
\affiliation{Panjab University, Chandigarh 160014, India}
\affiliation{Delhi University, Delhi-110 007, India}
\affiliation{Tata Institute of Fundamental Research, Mumbai-400 005, India}
\affiliation{University College Dublin, Dublin 4, Ireland}
\affiliation{Korea Detector Laboratory, Korea University, Seoul, 02841, Korea}
\affiliation{CINVESTAV, Mexico City 07360, Mexico}
\affiliation{Nikhef, Science Park, 1098 XG Amsterdam, the Netherlands}
\affiliation{Radboud University Nijmegen, 6525 AJ Nijmegen, the Netherlands}
\affiliation{Joint Institute for Nuclear Research, Dubna 141980, Russia}
\affiliation{Institute for Theoretical and Experimental Physics, Moscow 117259, Russia}
\affiliation{Moscow State University, Moscow 119991, Russia}
\affiliation{Institute for High Energy Physics, Protvino, Moscow region 142281, Russia}
\affiliation{Petersburg Nuclear Physics Institute, St. Petersburg 188300, Russia}
\affiliation{Instituci\'{o} Catalana de Recerca i Estudis Avan\c{c}ats (ICREA) and Institut de F\'{i}sica d'Altes Energies (IFAE), 08193 Bellaterra (Barcelona), Spain}
\affiliation{Uppsala University, 751 05 Uppsala, Sweden}
\affiliation{Taras Shevchenko National University of Kyiv, Kiev, 01601, Ukaine}
\affiliation{Lancaster University, Lancaster LA1 4YB, United Kingdom}
\affiliation{Imperial College London, London SW7 2AZ, United Kingdom}
\affiliation{The University of Manchester, Manchester M13 9PL, United Kingdom}
\affiliation{University of Arizona, Tucson, Arizona 85721, USA}
\affiliation{University of California Riverside, Riverside, California 92521, USA}
\affiliation{Florida State University, Tallahassee, Florida 32306, USA}
\affiliation{Fermi National Accelerator Laboratory, Batavia, Illinois 60510, USA}
\affiliation{University of Illinois at Chicago, Chicago, Illinois 60607, USA}
\affiliation{Northern Illinois University, DeKalb, Illinois 60115, USA}
\affiliation{Northwestern University, Evanston, Illinois 60208, USA}
\affiliation{Indiana University, Bloomington, Indiana 47405, USA}
\affiliation{Purdue University Calumet, Hammond, Indiana 46323, USA}
\affiliation{University of Notre Dame, Notre Dame, Indiana 46556, USA}
\affiliation{Iowa State University, Ames, Iowa 50011, USA}
\affiliation{University of Kansas, Lawrence, Kansas 66045, USA}
\affiliation{Louisiana Tech University, Ruston, Louisiana 71272, USA}
\affiliation{Northeastern University, Boston, Massachusetts 02115, USA}
\affiliation{University of Michigan, Ann Arbor, Michigan 48109, USA}
\affiliation{Michigan State University, East Lansing, Michigan 48824, USA}
\affiliation{University of Mississippi, University, Mississippi 38677, USA}
\affiliation{University of Nebraska, Lincoln, Nebraska 68588, USA}
\affiliation{Rutgers University, Piscataway, New Jersey 08855, USA}
\affiliation{Princeton University, Princeton, New Jersey 08544, USA}
\affiliation{State University of New York, Buffalo, New York 14260, USA}
\affiliation{University of Rochester, Rochester, New York 14627, USA}
\affiliation{State University of New York, Stony Brook, New York 11794, USA}
\affiliation{Brookhaven National Laboratory, Upton, New York 11973, USA}
\affiliation{Langston University, Langston, Oklahoma 73050, USA}
\affiliation{University of Oklahoma, Norman, Oklahoma 73019, USA}
\affiliation{Oklahoma State University, Stillwater, Oklahoma 74078, USA}
\affiliation{Oregon State University, Corvallis, Oregon 97331, USA}
\affiliation{Brown University, Providence, Rhode Island 02912, USA}
\affiliation{University of Texas, Arlington, Texas 76019, USA}
\affiliation{Southern Methodist University, Dallas, Texas 75275, USA}
\affiliation{Rice University, Houston, Texas 77005, USA}
\affiliation{University of Virginia, Charlottesville, Virginia 22904, USA}
\affiliation{University of Washington, Seattle, Washington 98195, USA}
\author{V.M.~Abazov} \affiliation{Joint Institute for Nuclear Research, Dubna 141980, Russia}
\author{B.~Abbott} \affiliation{University of Oklahoma, Norman, Oklahoma 73019, USA}
\author{B.S.~Acharya} \affiliation{Tata Institute of Fundamental Research, Mumbai-400 005, India}
\author{M.~Adams} \affiliation{University of Illinois at Chicago, Chicago, Illinois 60607, USA}
\author{T.~Adams} \affiliation{Florida State University, Tallahassee, Florida 32306, USA}
\author{J.P.~Agnew} \affiliation{The University of Manchester, Manchester M13 9PL, United Kingdom}
\author{G.D.~Alexeev} \affiliation{Joint Institute for Nuclear Research, Dubna 141980, Russia}
\author{G.~Alkhazov} \affiliation{Petersburg Nuclear Physics Institute, St. Petersburg 188300, Russia}
\author{A.~Alton$^{a}$} \affiliation{University of Michigan, Ann Arbor, Michigan 48109, USA}
\author{A.~Askew} \affiliation{Florida State University, Tallahassee, Florida 32306, USA}
% ...
%
%
\collaboration{The D0 Collaboration\footnote{with visitors from
$^{a}$Augustana College, Sioux Falls, SD 57197, USA,
$^{b}$The University of Liverpool, Liverpool L69 3BX, UK,
$^{c}$Deutshes Elektronen-Synchrotron (DESY), Notkestrasse 85, Germany,
$^{d}$CONACyT, M-03940 Mexico City, Mexico,
$^{e}$SLAC, Menlo Park, CA 94025, USA,
$^{f}$University College London, London WC1E 6BT, UK,
$^{g}$Centro de Investigacion en Computacion - IPN, CP 07738 Mexico City, Mexico,
$^{h}$Universidade Estadual Paulista, S\~ao Paulo, SP 01140, Brazil,
$^{i}$Karlsruher Institut f\"ur Technologie (KIT) - Steinbuch Centre for Computing (SCC),
D-76128 Karlsruhe, Germany,
$^{j}$Office of Science, U.S. Department of Energy, Washington, D.C. 20585, USA,
$^{k}$American Association for the Advancement of Science, Washington, D.C. 20005, USA,
$^{l}$Kiev Institute for Nuclear Research (KINR), Kyiv 03680, Ukraine,
$^{m}$University of Maryland, College Park, MD 20742, USA,
$^{n}$European Orgnaization for Nuclear Research (CERN), CH-1211 Geneva, Switzerland
and
$^{o}$Purdue University, West Lafayette, IN 47907, USA.
$^{\ddag}$Deceased.
}} \noaffiliation
\vskip 0.25cm

%% file: acknowledgement_APS_full_names.tex
%

We thank the staffs at Fermilab and collaborating institutions,
and acknowledge support from the
Department of Energy and National Science Foundation (United States of America);
Alternative Energies and Atomic Energy Commission and
National Center for Scientific Research/National Institute of Nuclear and Particle Physics  (France);
Ministry of Education and Science of the Russian Federation, 
National Research Center ``Kurchatov Institute" of the Russian Federation, and 
Russian Foundation for Basic Research  (Russia);
National Council for the Development of Science and Technology and
Carlos Chagas Filho Foundation for the Support of Research in the State of Rio de Janeiro (Brazil);
Department of Atomic Energy and Department of Science and Technology (India);
Administrative Department of Science, Technology and Innovation (Colombia);
National Council of Science and Technology (Mexico);
National Research Foundation of Korea (Korea);
Foundation for Fundamental Research on Matter (The Netherlands);
Science and Technology Facilities Council and The Royal Society (United Kingdom);
Ministry of Education, Youth and Sports (Czech Republic);
Bundesministerium f\"{u}r Bildung und Forschung (Federal Ministry of Education and Research) and 
Deutsche Forschungsgemeinschaft (German Research Foundation) (Germany);
Science Foundation Ireland (Ireland);
Swedish Research Council (Sweden);
China Academy of Sciences and National Natural Science Foundation of China (China);
and
Ministry of Education and Science of Ukraine (Ukraine).

%% file: Bs-Jpsif0.bbl
\begin{thebibliography}{99}

\bibitem{lhcb} R. Aaij {\sl et al}. (LHCb Collaboration), \textit{First observation of $B^0_s \rightarrow J/\psi f_0(980)$ decays}, Phys. Lett. B {\bf 698}, 115 (2011). 

\bibitem{bell} J. Li {\sl et al}. (Belle Collaboration), \textit{Observation of $B^0_s \rightarrow J/\psi f_0(980)$ and Evidence for $B^0_s \rightarrow J/\psi f_0(1370)$}, Phys. Rev. Lett. {\bf 106}, 121802 (2011). 

\bibitem{cdf} T. Aaltonen {\sl et al}. (CDF Collaboration), \textit{Measurement of branching ratio and $B_s^0$ lifetime in the decay $B^0_s \rightarrow J/\psi f_0(980)$ at CDF}, Phys. Rev. D {\bf 84}, 052012 (2011).

\bibitem{d0} V.M. Abazov {\sl et al}. (D0 Collaboration), \textit{Measurement of the relative branching ratio of $B^0_s \rightarrow J/\psi f_0(980)$ to $B^0_s \rightarrow J/\psi \phi$}, Phys. Rev. D {\bf 85}, 011103(R) (2012).

\bibitem{lhcblifetime} R. Aaij {\sl et al}. (LHCb Collaboration), \textit{Measurement of the $\bar{B}_s^0$ Effective Lifetime in the $J/\psi f_0(980)$ Final State}, Phys. Rev. Lett. {\bf 109}, 152002 (2012)

\bibitem{PDG2014} K.~A. Olive {\sl et al.} (Particle Data Group), \textit{Review of Particle Physics}, Chin. Phys. C {\bf 38}, 090001 (2014). 

\bibitem{lhcb1} R. Aaij et al. (LHCb Collaboration), \textit{Analysis of the resonant components in $\bar{B}^0_s \rightarrow J/\psi \pi^+ \pi^-$}, Phys. Rev. D {\bf 86}, 052006 (2012). 

\bibitem{lhcb2} R. Aaij et al. (LHCb Collaboration), \textit{Measurement of resonant and CP components in $\bar{B}^0_s \rightarrow J/\psi \pi^+ \pi^-$ decays}, Phys. Rev. D {\bf 89}, 092006 (2014). 

\bibitem{run2det} V.M. Abazov {\sl et al.}, \textit{The upgraded D0 detector}, Nucl.\ Instrum.\ Methods A {\bf 565}, 463 (2006).

\bibitem{etadef} $\eta=-\ln[\tan(\theta/2)]$, where $\theta$ is the polar angle with respect to the beamline.


\bibitem{pythia} T. Sj\"ostrand, S. Mrenna and P.Z. Skands, \textit{PYTHIA 6.4 Physics and Manual}, J. High Energy Phys. {\bf 05}, 026 (2006)

\bibitem{evtgen} D.G.  Lange,  \textit{The EvtGen    particle    decay    simulation   package},   Nucl.  Instrum.    Methods    in Phys. Res. A {\bf462}, 152 (2001); for details see \url{http://www.slac.stanford.edu/~lange/EvtGen}.

\bibitem{geant3} R. Brun {\sl et al.}, \textit{GEANT detector description and simulation tool}, CERN program library long writeup W, {\bf5013}, (1993).

\bibitem{pv} The 
primary vertex of the $p\bar{p}$ interaction is determined for each event 
using the average transverse position of the beam-collision point as a 
constraint.

\bibitem{punzi} G. Punzi, \textit{Comments on Likelihood fits with variable resolution}, arXiv physics/0401045 (2004).

\bibitem{cranmer} K. Cranmer, \textit{Kernel estimation in high-energy physics}, Comput. Phys. Comm. {\bf 136}, 3 (2001).

\bibitem {flatte} J.B. Gay {\em et al.}, \textit{Production and decay properties of the $\delta$(970) meson}, Phys. Lett. B {\bf 63}, 220 (1976). 

\bibitem {flatte2} S.~M. Flatt\'e, \textit{Coupled-channel analysis of the $\pi\eta$ and $K\bar{K}$ systems near $K\bar{K}$ threshold}, Phys. Lett. B {\bf 63}, 224 (1976).

\bibitem {flatte3} S.~M. Flatt\'e, \textit{On the nature of $0+$ mesons}, Phys. Lett. B {\bf 63}, 228 (1976).

\bibitem{PRL94-102001} V.M. Abazov {\sl et al.}, \textit{Measurement of the $\Lambda_b^0$ Lifetime in the Decay $\Lambda_b^0 \rightarrow J/\psi \Lambda^0$ with the D0 Detector}, Phys. Rev. Lett. {\bf 94}, 102001 (2005).



\end{thebibliography}
